\documentclass[a4paper,journal]{IEEEtran} 

\usepackage{color}
\usepackage{amsthm}
\usepackage{amsmath}
\usepackage{amssymb}
\usepackage{graphicx}
\usepackage[normalem]{ulem}
\usepackage{soul}

\makeatletter

\theoremstyle{plain}
\newtheorem{thm}{\protect\theoremname}
\theoremstyle{plain}
\newtheorem{lem}[thm]{\protect\lemmaname}

\makeatother

\providecommand{\lemmaname}{Lemma}
\providecommand{\theoremname}{Theorem}

\begin{document}

 \title{Precoding for PAPR Reduction in UW-OFDM }

\author{Morteza Rajabzadeh,~\IEEEmembership{Member,~IEEE,}
and~Heidi Steendam,~\IEEEmembership{Senior Member,~IEEE}
\thanks{M. Rajabzadeh is with the Electrical Engineering
Department, Quchan University of Technology, Quchan, Khorasan
Razavi, Iran (e-mail: m.rajabzadeh@qiet.ac.ir).
H. Steendam is with the Department of Telecommunications
and Information Processing TELIN/IMEC, Ghent University, Sint-Pietersnieuwstraat
41, 9000 Gent, Belgium (e-mail: Heidi.Steendam@ugent.be). 
The second author gratefully acknowledges the financial support from the Belgian Research Councils FWO and FNRS through the EOS grant 30452698, and the Flemish Government for the "Onderzoeksprogramma Artifici\"{e}le Intelligentie (AI) Vlaanderen" programme.   
}}
\maketitle
\begin{abstract}
Unique Word Orthogonal Frequency Division multiplexing (UW-OFDM) is a  variant of the multicarrier technique that has shown better bit error performance compared to cyclic prefixed (CP-)OFDM. Like other multicarrier techniques, UW-OFDM suffers from the high peak-to-average-power ratio (PAPR) problem, which is an obstacle ahead of practical implementation of UW-OFDM. In this paper, the generator matrix, which is inherent to UW-OFDM, is not only used to create the specific UW-OFDM system, but also to reduce the PAPR. To do so, a suitable optimization problem is developed, and solved using the Procrustes method. The main advantage of the proposed method is that the built-in generator matrix should be calculated only once for each system setting, and the PAPR reduction (PAPR-R) is achieved without any overhead redundancy. Simulation results are provided for evaluating the PAPR-R performance of the proposed method. 
\end{abstract}
\begin{IEEEkeywords}
Multicarrier systems, UW-OFDM, Peak to Average Power Ratio, optimization, precoding.
\end{IEEEkeywords}

\vspace{-15pt}
\section{Introduction}
The Unique Word (UW)-OFDM signaling scheme  \cite{UW-OFDM/NonSys/TSP2012/Huemer} is a multicarrier (MC) technique, for which the time-domain (TD) guard intervals are filled with an arbitrary known sequence -the unique word (UW)- instead of the random cyclic prefix (CP). The UW provides the same advantages as a CP: it makes the diagonalization of the channel matrix possible by the discrete Fourier transform (DFT) and circumvents intersymbol interference. On the other hand, as the UW is known, it can be used to estimate the channel or synchronize the signal \cite{UW-OFDM/PilotTones/TSP/2020}\cite{UW-SC-FDE/Synch/LComm/2018}\cite{UW-OFDM/MIMO/Fettweis/TWC/2020}. Furthermore, in distinction to CP-OFDM, the UW is already part of the DFT interval.  To make the insertion of the UW possible, a TD data block is generated that contains zeros at the positions of the UW by adding redundancy in the frequency domain using a generator matrix. This redundancy results in  several advantageous properties such as inherent diversity gain resulting in outstanding bit error rate (BER) performance \cite{UW-OFDM/Heidi/2016/theoretical}, and lower out-of-band radiation (OOBR) than for CP-OFDM \cite{UW-OFDM/PSD/Rajabzadeh/TCom/2017}. Yet, these advantageous properties come at some cost: to fully exploit the introduced diversity gain, UW-OFDM needs more sophisticated data detectors compared to conventional CP-OFDM \cite{UW-OFDM/NonSys/TSP2012/Huemer,UW-OFDM/DataEstimators/TSP/2011}. 
Considering practical implementation, all MC variants are to a greater or lesser extent susceptible to non-linear distortion \cite{OFDM/HPA/1999/ComL} because of the high peak-to-average-power ratio (PAPR). In contrast to CP-OFDM, where PAPR reduction (PAPR-R) has received a lot of attention , limited work is done for UW-OFDM. To the authors' best knowledge, the only study investigating the PAPR-R for UW-OFDM is \cite{PAPR/UW-OFDM/SLM/Huber/2012}, where the authors adopted the Selective Mapping (SLM) technique to reduce the PAPR of UW-OFDM. In \cite{UW-OFDM/PAPR/moaveni/2018/ICEE}, the UW-OFDM system parameters that affect its PAPR are studied without introducing any method to reduce the PAPR. 
In this paper, we propose a novel technique to reduce the PAPR in UW-OFDM. We employ the degrees of freedom that are available in the design of the generator matrix to make it operate as a PAPR-reducing precoder (PRP). By optimizing this precoder, we are able to lower the PAPR. We show by simulations that the PAPR-R of PRP is moderate compared to  two conventional PAPR-R techniques, i.e.  Partial Transmission Sequence (PTS) and SLM, however, it is achieved without reducing the data throughput, or increasing the required transmit energy or computational complexity. 

\vspace{-10pt}
\section{System Model\label{sec:System-Model}}
\vspace{-5pt}
\subsection{The UW-OFDM Signal}
\vspace{-3pt}
Consider a UW-OFDM system \cite{UW-OFDM/NonSys/TSP2012/Huemer} having $N$ subcarriers and a UW with length $N_u$ samples. To construct a data block with $N_u$ zeros in the TD, we need to add redundancy that is equivalent to $N_r \geq N_u$ symbols in the frequency domain. Further, similar to conventional OFDM systems, $N_z$ zero subcarriers are inserted at the DC carrier and at band edge carriers to serve as guard bands. This implies $N_d = N-N_r-N_z$ data symbols can be transmitted during a data block. We define $N_{dr} = N_d+N_r = N-N_z$ as the number of modulated subcarriers. The vector $\mathbf{d}\in\mathbb{C}^{N_{d}\times1}$ is the vector of $N_d$ zero-mean i.i.d. data symbols with $E\{\mathbf{d}\mathbf{d}^{H}\}=\sigma^2\mathbf{I}_{N_d}$, where $\mathbf{I}_M$ is the $M\times M$ identity matrix. The transmitter adds redundancy by applying this data vector $\mathbf{d}$ to the generator matrix $\mathbf{G}\in\mathbb{C}^{N_{dr}\times N_{d}}$ , and the resulting sequence is mapped on the subcarriers using the mapping matrix $\mathbf{B}\in\mathbb{R}^{N\times N_{dr}}$, which is a reduced version of the identity matrix in which the columns corresponding to the zero subcarriers are removed. Thus, the $N\times1$ frequency domain data vector is given as
$
\tilde{\mathbf{x}}=\mathbf{BG}\mathbf{d}.\label{eqn:FDdata_Sys-UW,general}
$
The vector $\tilde{\mathbf{x}}$ is converted to the TD with the $N\times N$ inverse DFT (IDFT) matrix $\mathbf{F}_{N}^{H}$, resulting in the TD data vector $\mathbf{x}'$ as
\begin{equation}
	\mathbf{x}'=\mathbf{F}_{N}^{H}\mathbf{BG}\mathbf{d}=\begin{bmatrix}\mathbf{x}_{d}\\
	\mathbf{0}_{\mathit{N_{u}}\times1}
	\end{bmatrix},\label{eq:TD,Bef-UW,general}
\end{equation}
where $\mathbf{x}_d \in \mathbb{C}^{(N-N_u)\times 1}$ is the non-zero part of the TD UW-OFDM data block. Then, the $N_{u}\times1$ UW vector $\mathbf{x}_{u}$ is added to the zero part of $\mathbf{x}'$ to form the TD data block to be transmitted over the channel:
\begin{equation}
\mathbf{x}=\mathbf{x}'+\begin{bmatrix}\mathbf{0}_{\mathit{(N-N_{u})}\times1}\\
\mathbf{\mathbf{x}}_{u}
\end{bmatrix}=\begin{bmatrix}\mathbf{x}_{d}\\
\mathbf{\mathbf{x}}_{u}
\end{bmatrix}.\label{eq:TD UW general form}
\end{equation}

\vspace{-18pt}
\subsection{Design of the Generator Matrix}\label{sub: II-A_G matrix}
To design the generator matrix \cite{UW-OFDM/NonSys/TSP2012/Huemer}  \cite{UW-OFDM/SlSupp/Rajabzadeh/EW/2014}, we consider the null space based approach  \cite{UW-OFDM/SlSupp/Rajabzadeh/EW/2014}: the matrix $\mathbf{F}_{N}^{H}\mathbf{B}$ is partitioned in two matrices, i.e. the matrices $\mathbf{A}\in\mathbb{C}^{(N-N_{u})\times N_{dr}}$ and $\mathbf{Q}\in\mathbb{C}^{N_{u}\times N_{dr}}$ that contain the $N-N_{u}$ upper and $N_{u}$ lower rows of $\mathbf{F}_{N}^{H}\mathbf{B}$, respectively: 
\begin{equation}
\mathbf{F}_{N}^{H}\mathbf{B}=\left[\begin{array}{c}
\mathbf{A}\\
\mathbf{Q}
\end{array}\right].\label{eq:Decomposition of FHB}
\end{equation}
In order to obtain the $N_{u}$ zeros at the output of IDFT, the generator matrix $\mathbf{G}$ must be selected in the null space of the matrix $\mathbf{Q}$. This is fulfilled by decomposing the matrix $\mathbf{G}$ as
\begin{equation}
\mathbf{G}=\mathbf{Y}\mathbf{C},\label{eq:Decomposed G}
\end{equation}
where $\mathbf{Y}=Null(\mathbf{Q})\in\mathbb{{C}}^{N_{dr}\times\left(N_{dr}-N_{u}\right)}$ is the matrix formed of $N_{dr}-N_{u}$ orthonormal null space basis vectors of $\mathbf{Q}$. This matrix is determined by the parameters from the UW-OFDM system, i.e., the number and positions of the zero subcarriers, $N_{u}$, $N_{r}$ and $N$, and is fixed, while the matrix $\mathbf{C}\in\mathbb{{C}}^{\left(N_{dr}-N_{u}\right)\times\left(N_{dr}-N_{r}\right)}$  can be selected freely. 

Besides the constraint that the matrix $\mathbf{G}$ has to be in the null space of the matrix $\mathbf{Q}$, the matrix $\mathbf{G}$ must satisfy some other conditions. In practical systems, often the transmitted energy per data symbol is normalized, leading to $trace\left[\mathbf{G}^{H}\mathbf{G}\right]=N_{d}$. Further, in \cite{UW-OFDM/NonSys/TSP2012/Huemer, UW-OFDM/Heidi/2016/theoretical}, it is shown that the condition 
\begin{equation}
\mathbf{G}^{H}\mathbf{G}=\mathbf{I}_{N_{d}},\label{eq:orthonormality const G}
\end{equation}
minimizes the mean-squared error of the linear minimum mean square error and the best linear unbiased estimator (BLUE) data detectors. Moreover, the condition (\ref{eq:orthonormality const G}) also turns out to be a sufficient condition to optimize the diversity gain \cite{UW-OFDM/Heidi/2016/theoretical}. Substituting the decomposition (\ref{eq:Decomposed G}) into the condition (\ref{eq:orthonormality const G}), and taking into account that $\mathbf{Y}^{H}\mathbf{Y=\mathbf{I}}_{(N_{dr}-N_{u})}$, we find that the matrix $\mathbf{C}$ must satisfy $\mathbf{C}^{H}\mathbf{C}=\mathbf{I}_{N_{d}}$. In the next section, we will propose a method to select the matrix $\mathbf{C}$, in order to minimize the PAPR.

\vspace{-10pt}
\section{PAPR Reduction by Precoding}\label{sec:PRP}
Let us rewrite the TD vector $\mathbf{x'}$ in (\ref{eq:TD,Bef-UW,general}) as follows:
\begin{equation}\label{eq:OVSed TD vector-summation}
	\mathbf{x'}=\mathbf{F}_{N}^{H}\mathbf{BG}\mathbf{d}=\mathbf{F}_{N}^{H}\mathbf{BY}\sum_{n=0}^{N_{d}-1}\mbox{\textbf{c}}_{n}d_{n}=		
		\left[\begin{array}{c} \mathbf{A}\\ \mathbf{Q} \end{array}\right]\mathbf{Y}\sum_{n=0}^{N_{d}-1}\mbox{\textbf{c}}_{n}d_{n},
\end{equation}
where $\textbf{c}_{n}$ is the $n$th column of $\mathbf{C}$. Hence, the TD data vector is obtained by the superposition
of the precoded data symbols $\textbf{c}_{n}d_{n}$ for $0\leq n\leq N_{d}-1$. The idea behind the proposed PRP scheme is to select $\mathbf{C}$ so that after the above-mentioned superposition, the $N_{d}$ data symbols occur at different time samples of the TD data vector $\mathbf{x'}$. In this way, the data are not constructively superposed, resulting in a reduction of the PAPR. As in the TD vector $\mathbf{x'}$ the last $N_u$ samples are zero, we will confine our attention to the $N-N_u$ first rows of $\mathbf{F}_{N}^{H}\mathbf{BY}$, and define the matrix $\mathbf{Z}=\mathbf{AY}\in\mathbb{C}^{(N-N_{u})\times (N_{dr}-N_u)}$. 

In order to obtain a TD sequence where the data symbols appear at different time instants, we need to select $\mathbf{C}$ so that  
\begin{equation}\label{eq:non-constrained PAPR-R relation}
	\mathbf{ZC}=\mathbf{D},
\end{equation}
where $\mathbf{D}$ is a $(N-N_{u})\times N_{d}$  matrix with
elements $\mathbf{D}_{k,k'}\in\{0,1\}$ and each column contains only
one '1' to make sure that each TD sample contains the contribution
of one data symbol. In this paper, we select the first column of $\mathbf{D}$ as $\{\mathbf{D}\}_{c=1}=[1,0,...,0]^{T}$,
and the $i$th column of $\mathbf{D}$ is obtained by shifting $\{\mathbf{D}\}_{c=1}$downward
$i$ times. As $(N-N_{u})\geq N_{d}$
 (since $N_d=N-N_r-N_z$ and $N_{r}\geq N_{u}$), i.e. the number of TD samples is larger than the number
of data symbols, in practice, some data symbols will occur at more than one TD
sample. In other words, since $\mathbf{Z}$ is not a square matrix,
to solve (\ref{eq:non-constrained PAPR-R relation}), we need to resort
to the following least squares problem: 
\begin{eqnarray} \underset{\mathbf{C}}{\min}\,\,\left\Vert \mathbf{ZC}-\mathbf{D}\right\Vert _{F}^{2},\end{eqnarray}
 where $\left\Vert .\right\Vert _{F}$ is the Frobenius norm operator. The solution of this unconstrained problem is: $\mathbf{C}=(\mathbf{Z}^H\mathbf{Z})^{-1}\mathbf{Z}^H \mathbf{D}$. However, to ensure that the obtained generator matrix shows good BER performance, we impose that $\mathbf{C}^H\mathbf{C}=\mathbf{I}_{N_d}$, resulting in the following constrained optimization problem:
\begin{eqnarray}\label{eq:constrained LS}
&&\underset{\mathbf{C}}{\min}\,\,\left\Vert \mathbf{ZC}-\mathbf{D}\right\Vert _{F}^{2}, \text{    s.t. } \mathbf{C}^{H}\mathbf{C}=\mathbf{I}_{N_d}.
\end{eqnarray}
This problem is a variant of the Procrustes problem \cite{Book/MatrixComputations/4th/golub/2013}. In general, the matrix $\mathbf{C}$ with  size $(N_{dr}-N_{u})\times (N_{dr}-N_r)$ is non-square. Although we considered the general case of $N_r\geq N_u$, in fact, $N_r=N_u$ redundancy is sufficient for the UW-OFDM system to work properly. In the following, we derive the solutions for two cases, i.e. $N_r=N_u$ and $N_r>N_u$.

\emph{I) Square $\mathbf{C} \ (N_r=N_u)$}: The problem (\ref{eq:constrained LS}) is a unitary Procrustes problem \cite{MatrixAnalysis/horn/1990}, where a unitary matrix must be found that most accurately transforms one matrix to another. The summary of the solution is given in the following lemma.

\begin{lem}
Let the matrices $\mathbf{U}$ and $\mathbf{V}$ be the left-hand and right-hand side unitary matrices of the singular value decomposition (SVD) of $\mathbf{D}^{H}\mathbf{Z}$, i.e. $\mathbf{D}^{H}\mathbf{Z}=\mathbf{U}\mathbf{\Sigma}\mathbf{V}^{H}$. The cost function (\ref{eq:constrained LS}) is minimized if $\mathbf{C}_{opt}=\mathbf{V}\mathbf{U}^{H}$.
\end{lem}
\begin{IEEEproof}
Let us rewrite the cost function of (\ref{eq:constrained LS}) as follows:
\begin{IEEEeqnarray}{lCr}  \label{eq:Proof_relation1}
\left\Vert \mathbf{ZC}-\mathbf{D}\right\Vert_F^2 =  &Trace[\mathbf{Z}\mathbf{C}\mathbf{C}^H\mathbf{Z}^H+\mathbf{D}^{H}\mathbf{D}] \nonumber \\ &-2 Real \lbrace Trace[\mathbf{D}^H\mathbf{ZC}] \rbrace.
\end{IEEEeqnarray} By assuming $\mathbf{C} \mathbf{C}^H=\mathbf{I}_{N_d}$, as the first term in  (\ref{eq:Proof_relation1}) is independent from $\mathbf{C}$, minimization of $\left\Vert \mathbf{Z}\mathbf{C}-\mathbf{D}\right\Vert _{F}^{2}$ is equivalent to maximization of $Real{\{Trace[\mathbf{D}^{H}\mathbf{Z}\mathbf{C}]\}}$. Decomposing the square $\mathbf{D}^{H}\mathbf{Z}$ using the SVD, i.e. $\mathbf{D}^{H}\mathbf{Z}=\mathbf{U}\mathbf{\Sigma}\mathbf{V}^{H}$, we obtain 
\begin{equation}
Trace[\mathbf{D}^{H}\mathbf{Z}\mathbf{C}]=Trace\left[\mathbf{U}\mathbf{\Sigma}\mathbf{V}^{H}\mathbf{C}\right]=Trace\left[\mathbf{\Sigma}\mathbf{K}\right],\label{eq:Trace value}
\end{equation}
where $\mathbf{K}=\mathbf{V}^H{\mathbf{C}}\mathbf{U}$ is a unitary matrix. As $Real\{Trace\left[\mathbf{\Sigma}\mathbf{K}\right]\}$ is maximized when $\mathbf{K}=\mathbf{I}$
\cite[7.4.5]{MatrixAnalysis/horn/1990}, the solution is: ${\mathbf{C}}_{opt}=\mathbf{V}\mathbf{U}^{H}$. Finally, the assumption $\mathbf{C} \mathbf{C}^H=\mathbf{I}$ is valid, because ${\mathbf{C}}, \mathbf{V} \text{ and } \mathbf{U}$ are square.
\end{IEEEproof} 
\emph{II) Non-Square $\mathbf{C} \ (N_r>N_u)$}: In this less-common case, the first right hand side term of (\ref{eq:Proof_relation1}) is a function of $\mathbf{C}$, and we cannot use Lemma 1. For the non-square $\mathbf{C}$, the problem (\ref{eq:constrained LS}) is known as an unbalanced Procrustes problem, which unfortunately, does not have a closed-form solution. Some   iterative algorithms are introduced for solving this problem, however, they have high complexities with different criteria for the convergence (for a recent study, please refer to \cite{opt/Procrustes/Unbalanced/SIAM/2020}). In this paper, we use the approximation $\mathbf{C}\mathbf{C}^H=\mathbf{I}$ for the non-square $\mathbf{C}$. Hence, we can use the solution of Lemma 1. We will study the applicability of the approximation by simulations.

The optimized $\mathbf{C}$ is
multiplied with $\mathbf{Y}$ to build the generator matrix $\mathbf{G}$
as in (\ref{eq:Decomposed G}). Note that the optimized $\mathbf{G}$ is independent of
the data symbols, because neither $\mathbf{Z}$ nor $\mathbf{D}$ depends on the transmitted data, but $\mathbf{G}$ only depends on the parameters $N$, $N_u$, $N_r$ and $N_d$ of the UW-OFDM.  So, we can precompute $\mathbf{G}$ for given system parameters, and use it while the UW-OFDM  parameters are not changed. This is one of the advantages of the PRP method over some of the classic PAPR-R techniques, especially PTS and SLM, in which
the selection of the best phase rotations should be
repeated for each UW-OFDM symbol, separately.  If we seek higher  levels of PAPR-R capability, PRP is flexible to be combined with  PTS or SLM, because the intrinsic structure of the UW-OFDM system is not changed in PRP. We call these combinations PRP-PTS and PRP-SLM.

\vspace{-10pt}

\section{Simulation Results\label{sec:simulations}}
In this section, the PAPR, BER and OOBR performances of the PRP are evaluated by computer simulations. The UW-OFDM system with no PAPR-R, and the UW-OFDM system with classic PTS and SLM techniques are considered as benchmarks. Similar to  most of the works on UW-OFDM, e.g.  \cite{UW-OFDM/NonSys/TSP2012/Huemer}, 
 \cite{UW-OFDM/PSD/Rajabzadeh/TCom/2017} and \cite{UW-OFDM/DataEstimators/TSP/2011}, we consider the system parameters  $N=64$, $N_d=36$, $N_z=12$ and $N_{r}=N_{u}=16$, unless stated otherwise.  The complete set of system parameters can be found in \cite[Table I]{UW-OFDM/PSD/Rajabzadeh/TCom/2017}.

\begin{figure}
\begin{centering}
\includegraphics[scale=0.42]{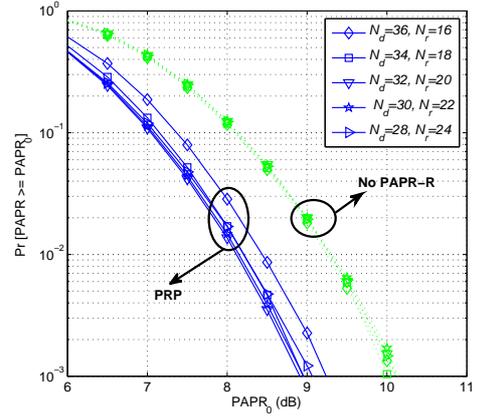}
\par\end{centering}
\caption{The CCDF of PAPR of UW-OFDM without PAPR-R and with the PRP method for different $N_r$ values, $N_d=52-N_r$, and fixed $N_u=16$. \label{fig:CCDF_UWOFDM_PRP_DiffNd}}
\vspace{-10pt}
\end{figure}

\emph{PAPR-R Performance}: To evaluate the PAPR, we use the complementary cumulative distribution function (CCDF) of the PAPR. The oversampling factor for having accurate PAPR results is $L=4$ \cite{PAPR/CP-ODFM/TR/Thesis/Tellado}.  In Fig. \ref{fig:CCDF_UWOFDM_PRP_DiffNd}, we show the PAPR for the PRP method for different values of $N_r$ (and so $N_d$) for a fixed $N_{dr} = N_d + N_r=$52.  It is seen that the PRP method is able to decrease the PAPR of UW-OFDM compared to the case where no PAPR-R scheme is deployed.  By decreasing $N_d$, the PAPR of UW-OFDM without PAPR-R does not change. Because although the number of transmitted data symbols ($N_d$) reduces, the number of  modulated subcarriers  $N_{dr}$ does not change. As the level of the PAPR is mainly determined by the number of subcarriers that is modulated, the PAPR will therefore not alter drastically.
 However, for UW-OFDM with PRP, decreasing $N_d$ results in lower PAPR values. This can be explained by having a closer look at the optimization problem in (\ref{eq:constrained LS}): when $N_d$ decreases, a lower number of transmitted data symbols ($N_d$) should be put on a fixed number of $(N-N_u)$ TD samples. Hence, the PRP is more successful to  decrease the superposition effect of these data symbols leading to lower PAPR values. Moreover, the results of this figure confirm that the approximation utilized in Section \ref{sec:PRP} for optimizing the non-square matrix $\mathbf{C}$ does not deteriorate the PRP scheme  performance for the case $N_r>N_u=16$. 

\begin{center}
\begin{figure}
\begin{centering}
\includegraphics[scale=0.42]{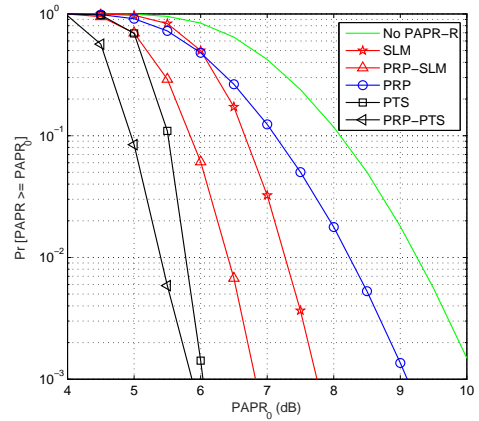}
\par\end{centering}
\caption{The CCDF of PAPR of the proposed PRP schemes for
UW-OFDM compared to the performance of SLM and PTS with $U=$ $V=4$.\label{fig:Fig2_CCDF_Precoding PAPR-R}}
\vspace{-15pt}
\end{figure}
\end{center}

The PAPR performance of PRP, SLM, PTS and the combinations PRP-SLM and PRP-PTS are shown in Fig. \ref{fig:Fig2_CCDF_Precoding PAPR-R}. For PTS, the number of sub-blocks is considered $V=4$, and for SLM, the number of implemented IDFT blocks is $U=4$. For both PTS and SLM, the phase rotations of each block are selected out of a set containing $W=4$ angles, i.e. the set $\{\pm1,\pm j\}$. Simulations not mentioned here for brevity show that increasing $U$ and $V$ results in a lower PAPR for both PTS and SLM as more combinations can be examined. However, when $U=V$, PTS is able to better reduce the PAPR than SLM, although this comes at the expense of a higher computational complexity to search for the optimal phase rotations. Indeed, PTS will select the transmitted sequence out of $W^V$ phase-rotated OFDM symbols, while in SLM, only $U$ phase-rotated symbols are compared to obtain the lowest PAPR.  It is also observed that the PAPR-R of the PRP is smaller than  PTS or SLM. Further, combining the PRP technique with PTS or SLM will reinforce the action of the PTS and SLM technique, and result in a considerably lower PAPR than achievable with PTS or SLM alone.
\vspace{-25pt}

\begin{figure}
\begin{centering}
\includegraphics[scale=0.40]{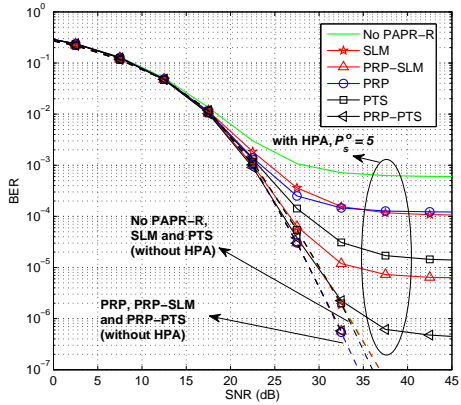}
\par\end{centering}
\vspace{-5pt}
\caption{BER performance of the PAPR-R schemes  without HPA nonlinearity (the dashed lines) and with HPA nonlinearity (the solid lines).\label{fig:Fig3_BER}}
\vspace{-15pt}
\end{figure}

\emph{BER Performance}: To study the BER performance, the signal data of the studied schemes are transmitted through a Rayleigh fading frequency selective channel with $L_{c}=$16 resolvable paths and an exponential power delay profile with  exponential decay factor 0.1. We assume the receiver has full channel state information, and use the BLUE data detector \cite{UW-OFDM/NonSys/TSP2012/Huemer}. In the transmitter, we consider the Rapp’s SSPA model \cite{PAPR/HPA/Rapp/1991}  to add the effect of the high power amplifier (HPA) nonlinearity to the simulations. Fig. \ref{fig:Fig3_BER} shows the BER performance of the schemes when the waveforms pass through the HPA nonlinearity with the knee factor $p=2$, and a clipping power level of $P^o_s$ = 5 dB above the mean power, which is a high clipping level. The results without HPA effect are also shown. For $P^o_s$ = 5 dB, because of high degree of clipping level, all the plots show an error floor at high signal-to-noise ratios (SNRs). In addition, the BER results confirm the results of Fig. \ref{fig:Fig2_CCDF_Precoding PAPR-R}: the lowest error floor belongs to the PRP-PTS, which has the best PAPR performance. 
When no HPA effect is considered, the high PAPR  does not affect the BER performance. It is seen that SLM and PTS do not alter the BER performance because neither the SLM nor the PTS method changes the basis structure of the UW-OFDM signal, as they only change the phases of the transmitted symbols, which are compensated at the receiver. On the other hand, in the PRP based methods, the BER performance slightly improves, e.g. the improvement is about 2 dB at BER=$10^{-7}$. Because although the constraint (\ref{eq:orthonormality const G}) guarantees to have full diversity at very high SNRs, the true diversity gain at moderate SNR levels will depend on the eigenvalues of the matrix $\mathbf{G}^H\mathbf{B}^H\mathbf{H}^H\mathbf{H}\mathbf{B}\mathbf{G}$, where $\mathbf{H}$ is the diagonal matrix with as diagonal elements the frequency response of the channel \cite{UW-OFDM/Heidi/2016/theoretical}. If one or more of these eigenvalues are much smaller than the other eigenvalues, this will cause a reduction of the diversity gain if the SNR is not sufficiently high. We observed in the proposed PRP technique that this difference between the eigenvalues reduces, causing the slight improvement in the BER performance for PRP based methods.

\begin{figure}
\begin{centering}
\includegraphics[scale=0.40]{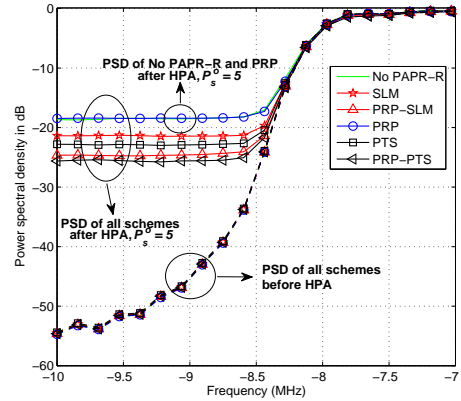}
\par\end{centering}
\vspace{-5pt}
\caption{The OOBR  of the PAPR-R schemes before passing through HPA nonlinearity (the dashed lines) and after HPA (the solid lines).\label{fig:Fig4_PSD}}
\vspace{-15pt}
\end{figure}

\emph{OOBR Performance}: The band edge of the power spectral density (PSD) of the studied schemes is illustrated in Fig. \ref{fig:Fig4_PSD}. Before the HPA, the OOBR of all schemes is the same with the same extent as extensively studied in \cite{UW-OFDM/PSD/Rajabzadeh/TCom/2017}. However, after the HPA non-linearity, the OOBR of all the schemes increases in accordance with the results of Fig. \ref{fig:Fig2_CCDF_Precoding PAPR-R}: with decreasing the PAPR-R performance, the OOBR also increases. The scheme with the least PAPR, i.e. PRP-PTS, shows the lowest OOBR increase. The OOBR of the PRP is almost the same as that of the UW-OFDM with no PAPR-R. 
\vspace{-6pt}

\vspace{-5pt}

\section{Conclusions}
\vspace{-5pt}
In this paper, a novel technique was proposed to reduce the PAPR of UW-OFDM signals by optimizing the generator matrix G. Proposed technique supports blind PAPR reduction alongside its main functionality, i.e. making zeros at the output of the IDFT in the UW-OFDM transmitter. Although it provides only about 1 dB PAPR improvement at CCDF=$10^{-3}$, the computational complexity can be ignored by using a look-up table before transmission. In fact, its sharing between the transmitter and the receiver is not considered as overhead redundancy of the PRP, because it needs to be shared for data detection purposes anyway. Further, PRP can be combined with several popular PAPR reduction schemes such as SLM and PTS for further improvement compared to either PTS, SLM or PRP itself.
\vspace{-10pt}
\bibliographystyle{IEEEtran}
\bibliography{IEEEfull,mybibfile}

\begin{thebibliography}{10}
\providecommand{\url}[1]{#1}
\csname url@samestyle\endcsname
\providecommand{\newblock}{\relax}
\providecommand{\bibinfo}[2]{#2}
\providecommand{\BIBentrySTDinterwordspacing}{\spaceskip=0pt\relax}
\providecommand{\BIBentryALTinterwordstretchfactor}{4}
\providecommand{\BIBentryALTinterwordspacing}{\spaceskip=\fontdimen2\font plus
\BIBentryALTinterwordstretchfactor\fontdimen3\font minus
  \fontdimen4\font\relax}
\providecommand{\BIBforeignlanguage}[2]{{%
\expandafter\ifx\csname l@#1\endcsname\relax
\typeout{** WARNING: IEEEtran.bst: No hyphenation pattern has been}%
\typeout{** loaded for the language `#1'. Using the pattern for}%
\typeout{** the default language instead.}%
\else
\language=\csname l@#1\endcsname
\fi
#2}}
\providecommand{\BIBdecl}{\relax}
\BIBdecl

\bibitem{UW-OFDM/NonSys/TSP2012/Huemer}
M.~Huemer, C.~Hofbauer, and J.~Huber, ``Non-systematic complex number {RS}
  coded {OFDM} by unique word prefix,'' \emph{IEEE Transactions on Signal
  Processing}, vol.~60, no.~1, pp. 285 --299, Jan. 2012.

\bibitem{UW-OFDM/PilotTones/TSP/2020}
C.~{Hofbauer}, W.~{Haselmayr}, H.~P. {Bernhard}, and M.~{Huemer}, ``On the
  inclusion and utilization of pilot tones in unique word {OFDM},'' \emph{IEEE
  Transactions on Signal Processing}, vol.~68, pp. 5504--5518, 2020.

\bibitem{UW-SC-FDE/Synch/LComm/2018}
J.~{Blumenstein} and M.~{Bobula}, ``Coarse time synchronization utilizing
  symmetric properties of zadoff–chu sequences,'' \emph{IEEE Communications
  Letters}, vol.~22, no.~5, pp. 1006--1009, May 2018.

\bibitem{UW-OFDM/MIMO/Fettweis/TWC/2020}
S.~{Ehsanfar}, M.~{Chafii}, and G.~P. {Fettweis}, ``On {UW}-based transmission
  for {MIMO} multi-carriers with spatial multiplexing,'' \emph{IEEE
  Transactions on Wireless Communications}, vol.~19, no.~9, pp. 5875--5890,
  2020.

\bibitem{UW-OFDM/Heidi/2016/theoretical}
H.~Steendam, ``Theoretical performance evaluation and optimization of
  {UW-OFDM},'' \emph{IEEE Trans. Commun.}, vol.~64, no.~4, pp. 1739--1750,
  2016.

\bibitem{UW-OFDM/PSD/Rajabzadeh/TCom/2017}
M.~Rajabzadeh and H.~Steendam, ``Power spectral analysis of {UW-OFDM}
  systems,'' \emph{IEEE Transactions on Communications}, vol.~66, no.~6, pp.
  2685--2695, June 2018.

\bibitem{UW-OFDM/DataEstimators/TSP/2011}
M.~Huemer, A.~Onic, and C.~Hofbauer, ``Classical and bayesian linear data
  estimators for unique word {OFDM},'' \emph{IEEE Transactions on Signal
  Processing}, vol.~59, no.~12, pp. 6073--6085, 2011.

\bibitem{OFDM/HPA/1999/ComL}
E.~Costa, M.~Midrio, and S.~Pupolin, ``Impact of amplifier nonlinearities on
  {OFDM} transmission system performance,'' \emph{IEEE Communications Letters},
  vol.~3, no.~2, pp. 37--39, February 1999.

\bibitem{PAPR/UW-OFDM/SLM/Huber/2012}
J.~B. Huber, J.~Rettelbach, M.~Seidl, and M.~Huemer, ``Signal shaping for
  unique-word {OFDM} by selected mapping,'' in \emph{18th European Wireless
  Conference}, April 2012, pp. 1--8.

\bibitem{UW-OFDM/PAPR/moaveni/2018/ICEE}
S.~M. Moaveni, M.~Rajabzadeh, and H.~Koshbin, ``A study on the {PAPR} of
  systematic {UW-OFDM},'' in \emph{Iranian Conference on Electrical Engineering
  (ICEE)}, May 2018, pp. 688--692.

\bibitem{UW-OFDM/SlSupp/Rajabzadeh/EW/2014}
M.~Rajabzadeh, H.~Khoshbin, and H.~Steendam, ``Sidelobe suppression for
  non-systematic coded {UW-OFDM} in cognitive radio networks,'' in \emph{20th
  European Wireless Conference}, May 2014, pp. 1--6.

\bibitem{Book/MatrixComputations/4th/golub/2013}
G.~H. Golub and C.~F. van Loan, \emph{Matrix Computations}, 4th~ed.\hskip 1em
  plus 0.5em minus 0.4em\relax Johns Hopkins University Press, 2013.

\bibitem{MatrixAnalysis/horn/1990}
R.~Horn and C.~Johnson, \emph{Matrix Analysis}.\hskip 1em plus 0.5em minus
  0.4em\relax Cambridge University Press, 1990.

\bibitem{opt/Procrustes/Unbalanced/SIAM/2020}
L.-H. Zhang, W.~H. Yang, C.~Shen, and J.~Ying, ``An eigenvalue-based method for
  the unbalanced {Procrustes} problem,'' \emph{SIAM Journal on Matrix Analysis
  and Applications}, vol.~41, no.~3, pp. 957--983, 2020.

\bibitem{PAPR/CP-ODFM/TR/Thesis/Tellado}
J.~Tellado, ``Peak to average power ratio reduction for multicarrier
  modulation,'' Ph.D. dissertation, University of Stanford, 1999.

\bibitem{PAPR/HPA/Rapp/1991}
C.~Rapp, ``Effects of {HPA-nonlinearity} on a {4-DPSK/OFDM-signal} for a
  digital sound broadcasting signal,'' in \emph{European Conference on
  Satellite Communications (ECSC)}, 1991, pp. 1245–--1253.

\end{thebibliography}
\end{document}